\documentclass[12pt]{article}
\pdfoutput=1
\usepackage{amsmath, amssymb}    
\usepackage{graphicx}
\usepackage{dcolumn}
\usepackage{bm}
\usepackage{hyperref}
\usepackage{color, xcolor}

\newcommand{\be}{\begin{equation} } 
\newcommand{\ee}{\end{equation} } 
\newcommand{\ba}{\begin{array} } 
\newcommand{\ea}{\end{array} } 
\newcommand{\bear}{\begin{eqnarray} } 
\newcommand{\eear}{\end{eqnarray} }

\begin{document}

\baselineskip=18pt \pagestyle{plain} \setcounter{page}{1}

\vspace*{-1cm}

\noindent \makebox[11.9cm][l]{\small \hspace*{-.2cm} }{\small Fermilab-PUB-26-0620-T}  \\  [-1mm]

\begin{center}

{\normalsize \Large  \bf   Axion-like particle at the 10 GeV scale: \\[3mm] Higgs decays to  wide jets and photons  
\\ [9mm]
}
\vspace*{0.3cm}

{\bf  Bogdan A. Dobrescu$^\diamond$ and Subhojit Roy$^\star$} 
\vspace{5mm}
\\
{\normalsize\small 
\it
$^\diamond$  Particle Theory Department, Fermilab, Batavia, IL 60510, USA     \\[2mm]
$^\star$ 
HEP Division, Argonne National Laboratory, Argonne, IL 60439, USA
\let\thefootnote\relax
\footnotetext[0]{bdob@fnal.gov, sroy@anl.gov}
}

\vspace{9mm}

{ \normalsize\small  August 18, 2026 }

\vspace{9mm}

\begin{abstract}  
If axion-like particles (ALPs) exist, they may have renormalizable couplings only to the Higgs boson ($h^0$)  or to fields beyond the Standard Model. An ALP $A_h$ lighter than about 60 GeV would allow the $h^0 \to A_h A_h$ decay, with a branching fraction determined by the explicit global symmetry breaking responsible for the main contributions to the $A_h$ mass ($M_A$). New heavy fields that carry color and electric charge, such as squarks, can mediate $A_h$ decays at one loop mostly into gluons, but also into photons. Exploring LHC sensitivity to $A_h$, we show that the  $h^0 \to A_h A_h \to (gg)(\gamma\gamma)$ channel leads to a diphoton resonance at $M_A$ whose production rate is consistent with a $3.5\sigma$ excess reported by a CMS search at $M_A \approx 13.6$ GeV. For $M_A$ of order 10 GeV, the $h^0 \to A_h A_h \to 4g$ cascade decay leads to two wide jets (each with 2-prong substructure) that form a resonance at 125 GeV.  
\end{abstract} 

\end{center}

\vfil

\thispagestyle{empty}  
  
\setcounter{page}{1}  
  
\vspace*{0.31cm}    
  
\newpage   
  
\tableofcontents
  
\vspace*{0.1cm}    
  
\baselineskip18pt   


\vspace*{0.6cm}   

\section{Introduction} 
\label{sec:intro}

Axion-like particles (ALPs) are predicted in various theories beyond the Standard Model (SM), and may have masses below the electroweak scale. 
Experimental sensitivity to ALPs with masses approximately in the $5 - 50$ GeV range is relatively poor~\cite{Bauer:2017ris}, given that production typically relies on LHC processes, and the SM  backgrounds are large in that mass range. Nevertheless, improved experimental methods (for example, scouting~\cite{CMS:2016ltu}, parked data~\cite{CMS:2024zhe}, etc.) 
employed by the ATLAS, CMS and LHCb Collaborations, as well as larger data sets, have increased the sensitivity to ALPs at the 10 GeV scale.

While ALPs are pseudoscalar particles associated with a shift symmetry that allows only derivative interactions, and thus, higher-dimensional operators, the ALP mass arises from explicit breaking of that shift symmetry. Consequently,  ALPs may also have non-derivative interactions, which furthermore can be renormalizable. However, the Lorentz and gauge structure of the SM  allows only the Higgs boson ($h^0$)  to have renormalizable interactions with an electroweak-singlet pseudoscalar. Therefore, 
precise studies of the Higgs boson properties may be sensitive to the existence of ALPs. 

In particular,  an ALP $A_h$ lighter than half the Higgs mass may provide nonstandard decay channels via $h^0 \to A_h A_h$~\cite{Dobrescu:2000jt}. 
In the presence of new heavy fields that carry color and electric charge, such as vectorlike quarks or squarks at the TeV scale, $A_h$ decays at one loop into two gluons, or with a much smaller branching fraction into two photons.
Here we explore possible discovery of $A_h$ in these channels. 

For an ALP of mass in the $10-20$ GeV range, the $h^0 \to A_h A_h \to (gg)(gg)$ cascade decay leads to two wide jets that form a resonance at 125 GeV that may be best probed in associated Higgs production. 
More promising is the  $h^0 \to A_h A_h \to (gg)(\gamma\gamma)$ channel, due to smaller backgrounds~\cite{Martin:2007dx}. The even cleaner $h^0 \to A_h A_h \to 4\gamma$ channel, searched for by CMS~\cite{CMS:2025mwx} and ATLAS~\cite{ATLAS:2023ian}, suffers from a highly-suppressed signal~\cite{Chang:2006bw} (see also~\cite{Brivio:2026qtx}).

A CMS search~\cite{CMS:2026zsp} for an ALP that is singly-produced via gluon fusion, and decays to a photon pair, has recently yielded an excess with a local significance of $3.5\sigma$  for a mass of 13.6 GeV. A similar ATLAS search~\cite{ATLAS:2022abz}, is not consistent with that excess under the assumption of a singly-produced ALP. Given that both the CMS and ATLAS searches mentioned here rely on a boosted photon pair, we propose the following alternative interpretation for the CMS excess. In the $h^0 \to A_h A_h \to (gg)(\gamma\gamma)$ process with the $A_h$ mass $M_A \approx 13.6$ GeV, the $\gamma\gamma$ pair has a momentum of $(M_h^2/4 - M_A^2)^{1/2} \approx 61$ GeV in the $h^0$ rest frame. Thus, the diphoton is automatically boosted and may naturally satisfy the CMS selection criteria, which are insensitive to the presence of the additional jets provided by the second $A_h$. 

Higgs decays into an ALP pair may be a first signal of an underlying origin of the Higgs sector, 
such as composite Higgs models~\cite{Dobrescu:1999gv} or the Next-to-Minimal Supersymmetric Model~\cite{Dobrescu:2000yn, Dermisek:2006wr, Datta:2022bvg}.
Although the focus of this paper is the ALP decays into photons or gluons, related ALP decays into $b$ quarks, taus, or other fermions have a long history of theoretical~\cite{Dobrescu:2000yn}-\cite{Bernreuther:2023uxh}  and experimental studies~\cite{ATLAS:2024vpj, CMS:2026mwx}.

In Section \ref{sec:model} we present a model based on a global symmetry associated with the ALP, and we analyze the spontaneous and explicit breakings of that symmetry.
In Section \ref{sec:LHC} we show that an ALP with $M_A \approx 13.6$ GeV that is produced in Higgs boson decays with a $h^0 \to A_h A_h$ branching fraction of about 10\% is consistent with the boosted diphoton results of both CMS and ATLAS. We then discuss additional features of this nonstandard Higgs decay that can increase the sensitivity of experimental searches.   
Independently of the current CMS excess, in Section \ref{sec:features}  we study possible LHC  signals arising from the Higgs boson decay to two wide jets, as a signal of pair produced ALPs.
 Our conclusions are summarized in Section \ref{sec:conclusions}.

\medskip  

\section{An axion-like particle coupled to the Higgs boson}     
\label{sec:model}   \setcounter{equation}{0}

We are studying an extension of the SM that includes  an axion-like particle $A_h$ (a pseudoscalar, {\it i.e.} a parity-odd spin-0 particle) coupled to the SM Higgs boson, and some heavy fields that carry color and electroweak charges.

\subsection{A complex scalar and a global symmetry}

Since $A_h$ is a pseudo-Nambu-Goldstone boson, it is useful to explicitly construct a renormalizable theory that exhibits the continuous global symmetry whose spontaneous and explicit  breakings lead to a mass for $A_h$ much smaller than the electroweak scale.
To that end, consider a gauge-singlet complex scalar $\phi$ whose potential is invariant under a $U(1)_\phi$ global symmetry:
\be
V(\phi) =   - M_\phi^2 \, \phi^\dagger \phi + \frac{\lambda_\phi}{2} \left( \phi^\dagger \phi  \right)^2  + \lambda_0 \, H^\dagger H \, \phi^\dagger \phi ~~,
\label{eq:scalar}
\ee
where $\lambda_\phi > 0$, $\lambda_0$ is a real dimensionless parameter, 
$M_\phi$ is a  mass parameter, and $H$ is the SM Higgs doublet. 
For $M_\phi^2 > \lambda_0 \, v_{_H}^2$, where $v_{_H} \! \approx 174$ GeV is the electroweak scale, the $\phi$ scalar acquires a
VEV:
\be
v_\phi = \frac{1}{\sqrt{\lambda_\phi } }   \left( M_\phi^2 -  \lambda_0 \, v_{_H}^2 \right)^{\! 1/2} ~~.
\label{eq:vphi}
\ee

In this case, the two degrees of freedom of $\phi$  are the pseudoscalar $A_h$ and a heavy real scalar $\varphi$, so that $\phi$  can be written as
\be
\phi = \left( v_\phi  + \frac{\varphi}{\sqrt{2}} \right) \, e^{i A_h /( \sqrt{2} \, v_\phi )}  ~~.
\label{eq:phi}
\ee
If the only Lagrangian terms involving $\phi$ are  a kinetic term and $- V(\phi)$, then $A_h$ remains exactly massless (being protected by a shift symmetry), and $\varphi$ acquires a tree-level mass $M_\varphi = \sqrt{ 2 \lambda_\phi } \, v_\phi$. The $\varphi$ mass  gets quadratically-divergent contributions at one loop, so we expect it to be of the same order or larger than the SM Higgs mass.

Let us now include a gauge-invariant Lagrangian term that explicitly breaks the global $U(1)_\phi$ symmetry and generates a coupling of $A_h$ to the SM Higgs boson. The simplest term of this type is 
\be
  \mu_{_{H \phi}}  \, \phi  \,  H^\dagger H   + {\rm H.c.}  ~~,
\label{eq:HHphi}
\ee
where the mass parameter satisfies $0 < \mu_{_{H \phi}} \ll v_\phi$ so that the minimization of the scalar potential does not significantly modify Eq.~(\ref{eq:vphi}).
The expansion of (\ref{eq:phi}) in powers of $A_h/v_\phi$, truncated after quartic couplings, gives
\be
\phi + \phi^\dagger = 2 \left( v_\phi  + \frac{\varphi}{\sqrt{2}} \right) \, \left( 1 - \frac{1}{4v_\phi^2} A_h^2 +  \frac{1}{96v_\phi^4} A_h^4 \right)  ~~.
\ee
Thus, the interaction (\ref{eq:HHphi}) includes the following important Lagrangian term:
\be
- \frac{\mu_{_{H \phi}}  }{2 v_\phi} \, H^\dagger H A_h A_h  ~~.
\label{eq:AAHH}
\ee
One of its implications is a trilinear coupling of the SM Higgs boson ($h^0$) to two $A_h$'s:
\be
- \frac{v_{_H}  \, \mu_{_{H \phi}} }{\sqrt{2}  \, v_\phi} \; h^0 A_h A_h   ~~.
\label{eq:hAAcoupling}
\ee 
This leads to a partial width of $h^0$ decay into ALPs given by 
\bear
&& \Gamma ( h^0  \to A_h A_h) \simeq  \frac{\mu_{_{H \phi}}^2 \, v_{_H}^2 }{16 \pi \, v_\phi^2 \, M_h}  \sqrt{1 - \frac{4 M_A^2}{M_h^2}}
\nonumber
\\ [2mm]
&& \hspace*{2.75cm}  \approx  10.5\% \, \left(\frac{\mu_{_{H \phi}}/ v_\phi }{0.01}\right)^{\! 2}  \, \Gamma ( h^0)_{\rm SM} ~~.
\label{eq:BRh}
\eear
The SM Higgs width is $\Gamma ( h^0)_{\rm SM} \approx 4.1$ MeV~\cite{ParticleDataGroup:2024cfk}.

\begin{figure}[t!]
\centerline{\hspace*{-.6cm}   \includegraphics[width=4.5cm, angle=0]{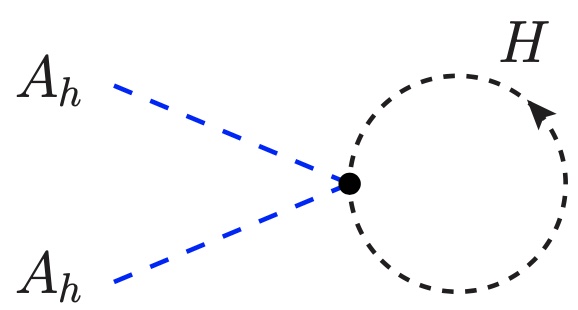}    } 
\vspace*{0.1cm}  
\caption{Quadratically divergent contribution to the squared-mass of the ALP $A_h$, due to the interaction (\ref{eq:AAHH}). The SM Higgs doublet is running in the loop.
\vspace*{2mm}
} 
\label{fig:Ahloop}
\end{figure}

Another implication of the term (\ref{eq:AAHH})  is that the ALP gets a tree-level contribution to its squared mass:
\be
(M_A^2)_{_1} = v_{_H}^2  \frac{\mu_{_{H \phi}}  }{ v_\phi}   ~~.
\label{eq:MA1}
\ee
In addition there are potentially large 1-loop contributions to the $A_h$ mass. 
To see that, note that the $H$ and $H^\dagger$ fields in (\ref{eq:AAHH}) can be connected such that a 1-loop contribution to $M_A^2$ is generated as in Figure~\ref{fig:Ahloop}.
This is a quadratically divergent contribution, which we estimate using an explicit momentum cut-off $\Lambda_H$:
\be
(M_A^2)_{_2}  \approx  \frac{\mu_{_{H \phi}} }{16 \pi^2 \, v_\phi}  \, \Lambda_H^2   ~~.
\label{eq:cutoff}
\ee
In Section \ref{sec:cut}, we will discuss a possible physical origin of the cut-off, and show that values for $\Lambda_H$ of order 1 TeV are natural. Thus, $(M_A^2)_2$ may be of the same order of magnitude as the tree-level contribution $(M_A^2)_1$.

\subsection{ALP coupled to heavy colored particles}

We introduce two scalars, $\tilde u_1$ and  $\tilde u_2$, which transform in the 
$(3,1,+2/3)$ representation of the SM gauge group. These color triplets (with masses in the TeV range) have an $U(1)_\phi$-invariant interaction with $\phi$:
\be
 \mu_{_{\tilde u \phi}}  \, \phi  \,  \tilde u_1^{\dagger}   \tilde u_2   + {\rm H.c.} 
\label{eq:uscalar2}
\ee
Any complex phase of the mass parameter $\mu_{_{\tilde u \phi}}$ can be absorbed by a redefinition of the colored scalar fields,
so we impose $\mu_{_{\tilde u \phi}} > 0$ without loss of generality.  The above Lagrangian term includes interactions of $\tilde u_1^{\dagger}  \tilde u_2$  with any number of $A_h$'s, as follows from the expansion of (\ref{eq:phi}).  The leading terms are given by
\be
 \mu_{_{\tilde u \phi}}  \left( v_\phi + \frac{ i}{ \sqrt{2} } \,  A_h - \frac{1}{4v_\phi} A_h A_h  \right)  \tilde u^{\dagger}_1  \tilde u_2 + {\rm H.c.} 
\label{eq:Ahu}
\ee 

Using the normalization where the global $U(1)_\phi$ charge of $\phi$ is +1, the $U(1)_\phi$ charges of $\tilde u_1$ and  $\tilde u_2$ can take any rational values that satisfy 
$x_1 - x_2 = 1$.  This ensures that (\ref{eq:uscalar2}) is invariant under $U(1)_\phi$. 
We also include a term in the Lagrangian that explicitly breaks the $U(1)_\phi$, for example the following mass mixing term:
\be
- m_{_{12}}^2  e^{i \beta_{12}} \,  \tilde  u_1^\dagger   \tilde u_2  + {\rm H.c.}  ~~,
\label{eq:umixing}
\ee
with mass parameter $m_{_{12}}  > 0$, and complex phase $\beta_{12} \in (0, 2\pi)$.  
The real part of the coefficient of the $ \tilde u_1^\dagger   \tilde u_2$ term then becomes 
\be
- m_{_{\rm R}}^2  =  - m_{_{12}}^2 \cos\beta_{12}+  \mu_{_{\tilde u \phi}}  v_\phi  ~~,
\ee
as it gets contributions from both (\ref{eq:umixing}) and (\ref{eq:Ahu}).  

\begin{figure}[t!]
\centerline{\hspace*{-0.1cm}   \includegraphics[width=7.2cm, angle=0]{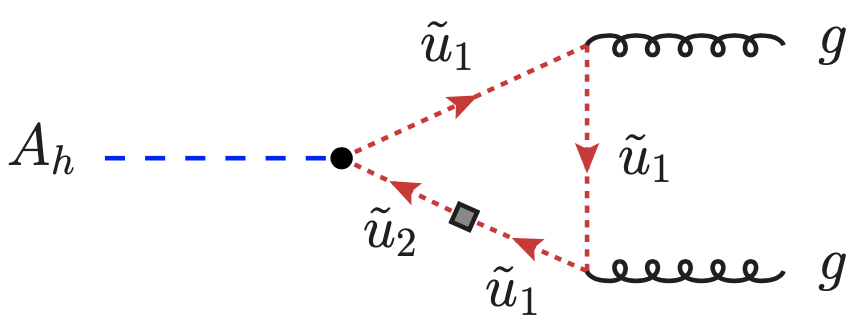}   \hspace*{1.3cm}    \includegraphics[width=5cm, angle=0]{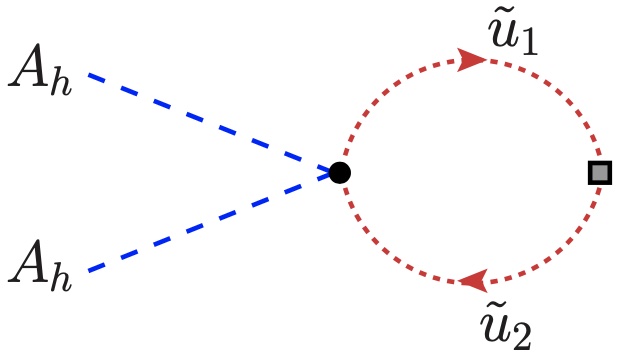}   } 
\vspace*{0.1cm}  
\caption{ One-loop consequences of the interaction (\ref{eq:uscalar2}), depicted as  $\bullet$, and of the (\ref{eq:umixing}) mixing, depicted as  {\scriptsize $   \textcolor{lightgray}{\blacksquare}  \hspace{-0.8em}  \textcolor{black}{\boldsymbol \square}$}. 
The color-triplet scalars  $\tilde u_1$ and  $\tilde u_2$ are running in the loop.
{\it Left:} Representative diagram responsible for the effective interaction of the ALP $A_h$ with gluons. {\it Right:} Logarithmically divergent contribution to the squared-mass of $A_h$. 
\vspace*{2mm}
} 
\label{fig:u12loop}
\end{figure}

The imaginary part of the coefficient  of $ \tilde u_1^\dagger   \tilde u_2$ in (\ref{eq:umixing}), together with the second term in (\ref{eq:Ahu}), leads to 
1-loop processes like the one shown in the left diagram of Figure~\ref{fig:u12loop}, which induces an effective coupling of the ALP to gluons. Similar diagrams have the mixing term  (\ref{eq:umixing})  inserted on the other colored-scalar lines, while additional diagrams with a $gg \tilde u_1^\dagger   \tilde u_1$ or $gg \tilde u_2^\dagger   \tilde u_2$ vertex are not shown. The same loops induce the effective coupling of $A_h$ to photons. Integrating out the colored scalars, we obtain the following 
effective couplings of $A_h$  to gluons and to photons:
\be
\frac{ \mu_{_{\tilde u \phi}}  \, m_{_{12}}^2 \sin\beta_{12}}{16 \pi  \sqrt{2} \, M_{\tilde u}^4 } 
f(M_{\tilde u_1} / M_{\tilde u_2}) 
\, A_h  \left(  \alpha_s G^{\mu\nu} G_{\mu\nu} + \frac{4}{3}  \alpha  F^{\mu\nu} F_{\mu\nu}  \right)  ~~,
\label{eq:gluons}
\ee
where $G^{\mu\nu}$ and $F^{\mu\nu}$ are the gluon and photon field strengths, respectively, and $M_{\tilde u} \approx (M_{\tilde u_1} + M_{\tilde u_2})/2$. The dimensionless function $f$, which depends on the ratio of colored-scalar masses, arises from the loop integration; its form is not relevant in what follows.
From (\ref{eq:gluons}) follows that the ALP has a partial width into gluons (related to the squark effects on $h^0\to gg$~\cite{Djouadi:1998az}) given by 
\be
\Gamma ( A_h  \to gg) =  \frac{\alpha_s^2 }{9216 \pi^3}   \left( f(M_{\tilde u_1} / M_{\tilde u_2}) \,  \frac{ \mu_{_{\tilde u \phi}}  \, m_{_{12}}^2 \sin\beta_{12} }{ M_{\tilde u_2}^4}   \right)^{\! 2}  M_A^3 ~~,
\ee
and the ratio of $\gamma\gamma$ and $gg$ widths is
\be
\frac{ \Gamma ( A_h  \to \gamma\gamma) }{\Gamma ( A_h  \to gg ) } = \frac{8\alpha^2 }{9 \alpha_s^2}   ~~.
\label{eq:ratio}
\ee
At the 10 GeV scale, $\alpha \approx 1/134$ and $ \alpha_s \approx 0.18$, so the branching fractions for $ A_h  \to \gamma\gamma$ 
and $ A_h  \to gg $ are approximately 0.15\% and 99\%, respectively. Three-body decay modes of $A_h$, such as $ggg$ or $gg\gamma$, are  suppressed and less interesting. 

The ratio (\ref{eq:ratio}) depends only on the color representations and electric charges of the particles in the loop, 
and is independent of their spin or $SU(2)_W$ representation. For example, replacing the colored scalars $\tilde u_1$ and $\tilde u_2$ with an up-type vectorlike quark (as in~\cite{Dobrescu:2000jt}) would not change the $ A_h$ branching fractions.
For a wide range of parameters, the $A_h$ decays are prompt at colliders, while single $A_h$ production is suppressed enough to make it very challenging to discover $A_h$ in the $pp \to A_h \to \gamma\gamma$ mode. 

The last term in (\ref{eq:Ahu}) in conjunction with the mixing (\ref{eq:umixing}) leads to the 
1-loop contribution to the $A_h$ squared mass shown in the right diagram of Figure~\ref{fig:u12loop}. That 
is a logarithmically-divergent contribution approximately given by 
\be
(M_A^2)_{_3}  \approx  - \frac{ 3 \, \mu_{_{\tilde u \phi}} m_{_{\rm R}}^2 }{16 \pi^2 \, v_\phi}  \, \ln \! \left(\frac{\Lambda_{\tilde u}}{M_{\tilde u}}\right)   ~~,
\label{eq:MAloop}
\ee
where $\Lambda_{\tilde u}$ is the momentum cut-off for the logarithmically-divergent integral, and for simplicity we assumed an approximate mass degeneracy between $\tilde u_1$ and $\tilde u_2$. 
Note that $\Lambda_{\tilde u}$ is generically different than the $\Lambda_H$ cut-off introduced in (\ref{eq:cutoff}).
As there are three contributions to the squared-mass of $A_h$ [see (\ref{eq:MA1}), (\ref{eq:cutoff}), (\ref{eq:MAloop})], 
$M_A^2 \approx (M_A^2)_{_1}  + (M_A^2)_{_2} + (M_A^2)_{_3}$, and we obtain the following formula for the ALP mass:
\be
M_A \approx  \sqrt{ \frac{\mu_{_{H \phi}} }{ v_\phi} } \left( v_{_H}^2 + \frac{\Lambda_H^2}{16 \pi^2} -  \frac{3 \,  \mu_{_{\tilde u \phi}} m_{_{\rm R}}^2 } {16 \pi^2  \mu_{_{H \phi}}}  
\, \ln \!\left(\Lambda_{\tilde u}/M_{\tilde u}\right)  \right)^{\! 1/2} ~~.
\ee
To have a sense of the range of parameters that yield $M_A$ at the 10 GeV scale, assume $\mu_{_{H \phi}} / v_\phi \approx 10^{-2}$, 
$\Lambda_H = O(1 \, {\rm TeV})$, $m_{_{\rm R}}/ v_{_H} = O(1)$, $\Lambda_{\tilde u}/M_{\tilde u} =  O(10^2)$. In that case $M_A$ decreases  monotonically from 18 GeV to 10 GeV when $\mu_{_{\tilde u \phi}} / \mu_{_{H \phi}}$ increases from 1 to 10.

\subsection{Physical origin of the Higgs loop cut-off}
\label{sec:cut} 

The momentum cut-off from (\ref{eq:cutoff}) may arise from perturbative new physics at the TeV scale.
Here we present a simple physical origin for this cut-off. 
The quadratically  divergent contribution to $M_A^2$ due to the SM Higgs doublet loop can be turned into a logarithmically divergent one 
in the presence of a second Higgs doublet, $H'$, with a large positive squared mass, $M^2_{\! _{H'}}$. To see that, first replace the term (\ref{eq:HHphi})  by 
the following $U(1)_\phi$ symmetric term:
\be
 - \mu_{_{H\phi}}^\prime  \, \phi  \,  H^\dagger H'   + {\rm H.c.}  ~~
\label{eq:3scalarsH}
\ee
Then, include a small $U(1)_\phi$ breaking term that mixes the two doublets:
\be
- m^2_{_{H\phi}}  \,  H^\dagger H'   + {\rm H.c.}  ~~,
\label{eq:2scalarsH}
\ee
where $m^2_{_{H\phi}} $ and $\mu_{_{H\phi}}^\prime$ are chosen such that after $H'$ is integrated out one recovers (\ref{eq:HHphi}), namely
$\mu_{_{H\phi}}^\prime m^2_{_{H\phi}}  = \mu_{_{H\phi}} M_{\! _{H'}}^2 $.  
Comparing  the ensuing 1-loop contribution to the $A_h$ squared mass  with (\ref{eq:cutoff}) gives
\be
\Lambda_H^2 =  4 M_{\! _{H'}}^2  \ln \left(\frac{\Lambda_H'}{M_{\! _{H'}}}\right)   ~~,
\ee 
where $\Lambda_H'$ is a physical momentum cut-off, presumably much above the TeV scale.
For $\Lambda_H'/M_{\!  _{H'}} = O(10)$ and $M_{\!  _{H'}} \approx 0.4$ TeV, the quadratic divergence is cut off at $\Lambda_H \approx 1.2$~TeV.

\bigskip

\section{LHC signals of $A_h$}
\label{sec:LHC}  \setcounter{equation}{0}

Single production of the ALP $A_h$ at the LHC is possible through gluon fusion, which is the inverse of the process shown in the left-hand diagram of Figure~\ref{fig:u12loop}. For a large range of parameters, however, the cross section for $gg \to A_h$ is too suppressed to allow discovery. Thus, pair production of $A_h$
in Higgs boson decays may offer the best probe for this ALP. 

SM Higgs boson production via gluon fusion in Run 2 of the LHC ($\sqrt{s} = 13$ TeV and integrated luminosity $L_{\rm int} \approx 139$ fb$^{-1}$) has a cross section of 48.6 pb with an uncertainty of about 7\%, based on N$^3$LO QCD for large $m_t$~\cite{Anastasiou:2016cez}. Including the other Higgs production processes (vector-boson fusion, $W/Z$ associated production, etc.) increases the cross section to approximately 56 pb~\cite{ParticleDataGroup:2024cfk}.
This implies that approximately   $8\times 10^6$ Higgs bosons have been produced in Run 2. 

That number increased by a factor of about 2.5 in Run 3  ($\sqrt{s} = 13.6$ TeV, $L_{\rm int} \approx 326$ fb$^{-1}$), mostly due to the larger integrated luminosity, but also due to an increase of roughly 8\% in the cross sections. Thus, if the $h^0 \to A_h A_h$ decay is allowed and its branching fraction is at the percent level, then more than $3 \times 10^5$ $A_h$ pairs have already been produced.

\medskip

\subsection{Kinematic features of the $h^0 \rightarrow A_h A_h$ signals} 
\label{sec:features}

It is useful to analyze the characteristic kinematic features of our signal.
The Higgs boson decay into two ALPs, $h^0 \rightarrow A_h A_h$, implies that in the Higgs rest frame each $A_h$ carries an energy $E_{A_h} = M_h / 2 \approx 62.5$ GeV. The corresponding Lorentz boost factor is 
\begin{equation}
\gamma_{A_h} \simeq \frac{E_{A_h}}{M_A} \approx   6.3   \left( \frac{10 \; {\rm GeV}}{M_A}  \right) \, \, .
\end{equation}

As a result, the decay products of $A_h$ with mass near the 10 GeV scale are expected to be
moderately collimated. An estimate for the angular separation of
two approximately massless decay products from a boosted resonance is
\begin{equation}
\Delta R \simeq  \frac{2M_A}{p_{\rm T}(A_h)}  ~~.
\end{equation}
The characteristic scale for the transverse momentum of $A_h$, 
$p_{\rm T}(A_h)\sim  (M_h^2/4 - M_A^2)^{1/2}$,  implies
\begin{equation}
\Delta R_{\gamma\gamma}  \simeq  \Delta R_{gg}  \approx  0.32   \left( \frac{M_A} {10 \; {\rm GeV}} \right) \left( 1 - 0.026 \,  \frac{M_A^2} {(10 \; {\rm GeV})^2  } \right)^{\! -1/2} ~.
\label{eq:deltaRestimate}
\end{equation}
For a typical jet clustering of cone $R=0.4$, the two gluon jets originating from an $A_h$ decay are likely to merge into a wide jet, labelled $J_{gg}$, which has a 2-prong substructure. 
By contrast, photon reconstruction in  the electromagnetic calorimeter benefits from a much finer cone, so that 
the two photons originating from the other $A_h$ decay, while relatively collimated, are sufficiently separated to be reconstructed as distinct objects in a large fraction of events. Consequently, requirements such as $\Delta R_{\gamma\gamma}>0.2$ or 0.3, employed by the CMS and ATLAS low-mass diphoton searches~\cite{CMS:2026zsp,ATLAS:2022abz}, should retain a substantial fraction of the signal.

 \begin{figure}[b]
    \centering
    \vspace*{4mm}
    \includegraphics[width=0.5\textwidth]{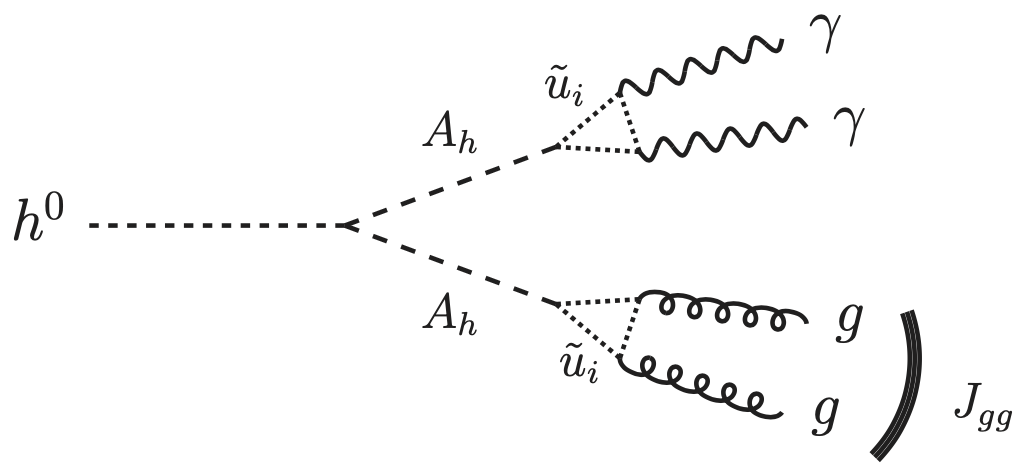}
    \caption{Nonstandard decay of the Higgs boson via a pair of $A_h$ ALPs. In the final state there are two photons and a 
    wide jet with 2-prong substructure:     $h^0 \to A_hA_h \to (\gamma\gamma) J_{gg}$.   }
    \label{fig:gammasgg}
\end{figure}

The above estimates are intended only as order-of-magnitude guidance.
In realistic proton-proton collisions, the Higgs boson is produced with
a non-trivial transverse-momentum spectrum, leading to broad
distributions of $A_h$ boosts and angular separations. Additional
effects from parton showering, hadronization, detector resolution,
photon reconstruction, and event selection further modify the observed
distributions. Nevertheless, the simple estimates above capture the
essential features of the signals, and provide a qualitative
understanding of why the topology considered here satisfies the boosted diphoton selections employed by
both CMS and ATLAS, while leading to wide jets with 2-prong substructure in the case of gluonic decays of $A_h$. 
In Section~\ref{sec:gammas} we perform a detailed detector-level study
of the corresponding experimental analyses. 

\begin{figure}[t]
    \centering
    \includegraphics[width=0.47\textwidth]{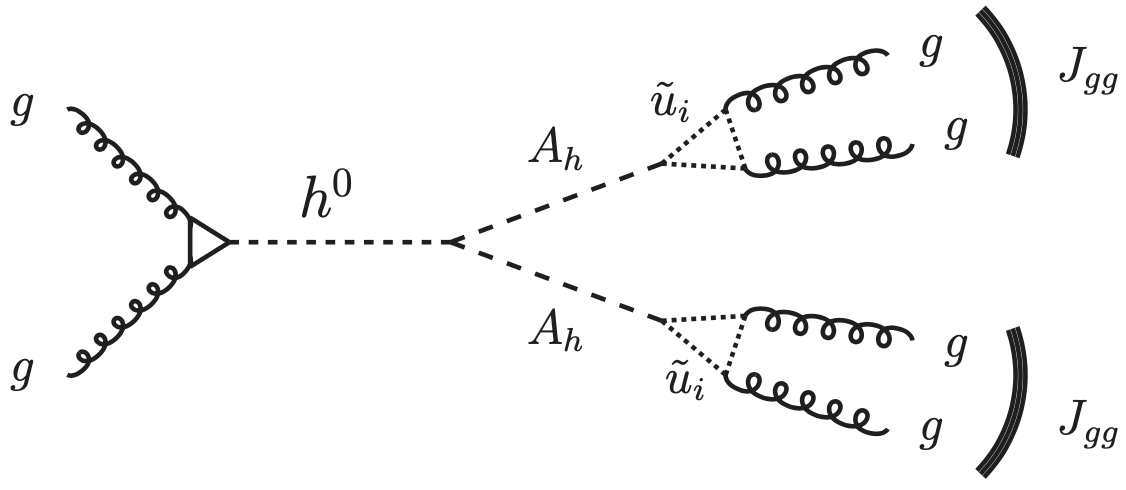}
    \hfill
    \includegraphics[width=0.47\textwidth]{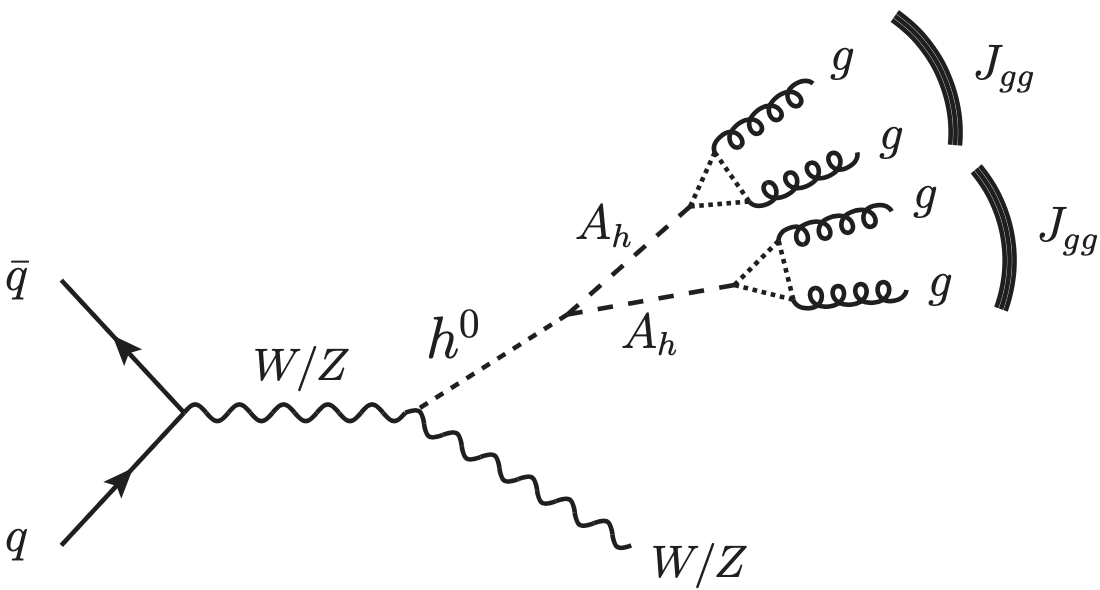}
    \caption{
      SM Higgs production through gluon fusion (left diagram) or associated production (right diagram),  followed by the nonstandard hadronic decay of $h^0$ 
      into two wide jets ($J_{gg}$), each with 2-prong substructure. When the final-state $W$ or $Z$ boson has a large $p_{\rm T}$, the two $J_{gg}$
      likely merge into a wider jet with a (2+2)-prong substructure, allowing further background suppression.
    }
    \label{fig:JJ}
\end{figure}

Given the $A_h  \to \gamma\gamma$ branching fraction of $10^{-3}$ implied by Eq.~(\ref{eq:ratio}), the rate of the clean 
process $pp \to h^0 \to A_h A_h  \to 4\gamma$ is too small to allow discovery of $A_h$.  
The related process where one $A_h$ decays to photons and the other to gluons (see  Figure~\ref{fig:gammasgg}), is the most promising one, as discussed in Section~\ref{sec:gammas}. 

Let us analyze briefly here the purely hadronic process $pp \to h^0 \to A_h A_h  \to 4g$, which among the new physics signals of the model presented in Section~\ref{sec:model}  has by far the largest rate. As the two gluon jets from an $A_h$ decay are likely to merge into a wide jet, the most relevant nonstandard Higgs decay is $h^0 \to A_h A_h  \to J_{gg}J_{gg}$, where each wide jet $J_{gg}$ has a 2-prong substructure. Figure~\ref{fig:JJ} shows this decay after Higgs boson production via gluon fusion (left-hand diagram) or in association with a $W/Z$ boson (right-hand diagram).

So far, no search has been performed for the Higgs boson decay into two wide jets. We urge the ATLAS and CMS Collaborations to consider searches of this type, despite the large backgrounds that would likely need to be estimated from the data. It is noteworthy that if the up-type colored scalars from the model presented in Section \ref{sec:model} are replaced by down-type ones, {\it i.e.} with electric charge $-1/3$, then the $A_h \to \gamma\gamma$ branching fraction is further suppressed by a factor of 16, leaving $h^0 \to A_h A_h  \to 4g$ the main discovery mode.

The large QCD backgrounds to the nonstandard hadronic Higgs processes can be suppressed by taking advantage of several kinematic features of the signals.
Each wide jet $J_{gg}$ has an invariant-mass distribution peaked near $M_A$, while invariant mass  of $J_{gg}J_{gg}$ has a broad distribution peaked near 125 GeV. 
Moreover, when the gluon fusion process includes initial-state radiation of high enough $p_{\rm T}$ (a special type of signal with smaller cross section), the Higgs boson itself is boosted so that the two $J_{gg}$ wide jets are likely to form a single wider jet with a peculiar (2+2)-prong substructure.

The same boosted $h^0$ topology occurs in a larger fraction of events in the case of associated production (even without initial-state radiation)
when the $W$ or $Z$ boson in the final state has a large boost. 
Furthermore, leptonic decays of the $W$ or $Z$ boson allow further suppression of the SM backgrounds. Hence, although associated Higgs production has a rate smaller by an order of magnitude than gluon fusion, it may offer the best channels for probing the nonstandard hadronic decays of $h^0$. 

Other Higgs production channels can also be included to improve the sensitivity of searches for such $h^0$ decays. Vector-boson fusion implies two forward jets in addition to the $J_{gg}J_{gg}$ system. Despite a smaller cross section~\cite{ParticleDataGroup:2024cfk}, $t\bar t h^0$ production offers several handles to reduce the backgrounds to the nonstandard  hadronic Higgs decay. 

\medskip

\subsection{Higgs decay to $A_h A_h \to  (\gamma\gamma)(gg)$}
\label{sec:gammas} 

We now investigate the collider phenomenology of $A_h$ through the process
\begin{equation}
pp \to h^0 \to A_h A_h  \to (\gamma\gamma) (gg)   ~~.
\end{equation}
This gives rise to a characteristic diphoton-plus-jets topology, where the two photons,  and separately the two gluon jets, form resonances at $M_A$.  
No dedicated search has yet been performed for this final state. However, both the 
ATLAS~\cite{ATLAS:2022abz} and CMS~\cite{CMS:2026zsp} Collaborations have performed inclusive searches for diphoton resonances in the $10 - 70$ GeV mass range, relying on boosted topologies that reduce the backgrounds. Those searches are sensitive to the $\gamma\gamma+$jets final state, and the largest excess ($3.5\sigma$) was reported at a diphoton invariant mass $m_{\gamma\gamma}\simeq 13.6~{\rm GeV}$ by CMS.  At the same mass ATLAS had a much smaller excess, with local significance of about $1.8\sigma$, based on a different set of selection criteria. 
Motivated by these observations, we investigate in this Section whether an ALP with a mass $M_A \approx13.6$ GeV can be efficiently
selected by the existing CMS and ATLAS analyses, and then study the
corresponding phenomenological implications.
%

\subsubsection{Sensitivity of CMS and ATLAS low-mass diphoton searches}
\label{sec:analysis}

Although the ATLAS~\cite{ATLAS:2022abz} and CMS~\cite{CMS:2026zsp} dedicated searches for
low-mass diphoton resonances were not designed for the process
$h^0\rightarrow A_h A_h \to (\gamma\gamma) J_{gg}$,
the characteristic momentum scale of the diphoton system arising from this process  naturally populates the boosted region targeted by those searches.  Here we present the features of the CMS and ATLAS searches implemented in our simulations, and then we compute the efficiencies of our nonstandard Higgs decay signal. 

The CMS search~\cite{CMS:2026zsp} employed a dedicated low-mass diphoton trigger together
with a neural-network classifier trained on photon-identification
variables, diphoton kinematics, mass-resolution observables, and
vertex-reconstruction information. The signal region is defined by a requirement on the neural-network output score. In contrast,
the ATLAS search~\cite{ATLAS:2022abz} exploited the kinematics of boosted diphoton systems to
extend sensitivity to masses down to 10 GeV while
maintaining control over the rapidly falling background distribution.
Although the experimental implementations differ, both analyses were designed to select energetic and isolated photon pairs originating from narrow resonances.

For our analysis, we have implemented the model described by (\ref{eq:hAAcoupling}) and  (\ref{eq:gluons})
in \texttt{FeynRules}~\cite{Alloul:2013bka} and generated the corresponding UFO model files~\cite{Darme:2023jdn}. The latter are interfaced with \texttt{MadGraph5\_aMC@NLO}~\cite{Alwall:2014hca} to compute matrix
elements and generate parton-level events, using  the PDF set NNPDF23\_nlo\_as\_0119\_qed~\cite{Ball:2013hta}. 
Decays of unstable particles, together with parton showering and
hadronization, are simulated using \texttt{PYTHIA8}~\cite{Sjostrand:2007gs}. Finally, detector responses are taken into account using {\tt DELPHES}~\cite{deFavereau:2013fsa} running with the 
CMS and ATLAS detector cards with some modifications in the photon isolation criteria, which are discussed below.

To estimate the acceptance of these searches for our signal, we perform analysis using the ROOT framework~\cite{Brun:1997pa}.
The main ingredients entering our implementations of the CMS~\cite{CMS:2026zsp} and ATLAS~\cite{ATLAS:2022abz} analyses are summarized below.

\noindent
{\it 1. Photon reconstruction and fiducial acceptance.}  
CMS requires two photons within the ECAL acceptance,
$|\eta|<2.5$, excluding the transition region  $1.44<|\eta|<1.57$. ATLAS requires photons satisfying
$|\eta|<2.37$ while excluding the calorimeter transition region $1.37<|\eta|<1.52$,
together with tight photon-identification requirements based on electromagnetic shower-shape observables.

\noindent
{\it 2. Trigger and photon transverse momentum requirements.}
A key feature of the CMS analysis is the dedicated low-mass diphoton
trigger requiring photons with transverse momenta
\begin{equation}
p_{\rm T}^{\gamma_1}>30~{\rm GeV} ~,
\qquad
p_{\rm T}^{\gamma_2}>18~{\rm GeV} ~,
\label{eq:pTgammaCMS}
\end{equation}
together with the scaled requirements
\begin{equation}
\frac{p_{\rm T}^{\gamma_1}}{m_{\gamma\gamma}}>0.47  ~,
\qquad
\frac{p_{\rm T}^{\gamma_2}}{m_{\gamma\gamma}}>0.28  ~.
\label{eq:pTgammaCMS2}
\end{equation}
While the ATLAS search similarly exploits boosted diphoton kinematics,
\begin{equation}
p_{\rm T}^{\gamma_1} , \; p_{\rm T}^{\gamma_2}> 22~{\rm GeV} ~,
\label{eq:pTgammaATL}
\end{equation}
 it does not employ scaled photon-momentum requirements. 

\noindent
{\it 3. Diphoton invariant-mass region.}
Both analyses search for narrow resonances in the low-mass diphoton
spectrum. In our implementation, we adopt 
 $10 < m_{\gamma\gamma} < 70~{\rm GeV}$
for the CMS-inspired selection and
$9 < m_{\gamma\gamma} < 77~{\rm GeV}$ 
for the ATLAS-inspired selection, corresponding to the ranges used in
the respective experimental analyses.

\noindent
{\it 4.  Photon isolation and reconstruction effects.}
Both searches set calorimeter- and track-based photon-isolation
requirements. In our {\tt DELPHES} implementation, we adopt isolation-cone
sizes consistent with the experimental analyses, namely
$\Delta R=0.3$ for the CMS card and
$\Delta R=0.2$ for the ATLAS card. However, we choose a 
large value of the isolation parameter ${\tt PTRatioMax}$, so that 
the isolation algorithm does not reject events due to
nearby energy deposits. This allows us to retain the full signal sample
and study the impact of photon collimation.
Since an issue for our signal is whether the two photons from
$A_h \rightarrow \gamma\gamma$ can be reconstructed as distinct objects,
we impose photon-separation requirements,
$ \Delta R_{\gamma\gamma}>0.3$ and 0.2
for the CMS- and ATLAS-inspired selections, respectively.
These selections provide an approximation of
the photon-reconstruction performance relevant for our signal,
rather than a detailed implementation of the experimental isolation
requirements.

\noindent
{\it 5.  Boosted diphoton topology.}
A requirement of the ATLAS analysis is 
\begin{equation}
p_{\rm T}^{\gamma\gamma}>50~{\rm GeV} ~~,
\label{eq:50}
\end{equation}
which suppresses trigger turn-on effects and improves the
description of the low-mass background spectrum. Although CMS does not
impose such a cut, their neural-network classifier strongly favors events with a boosted diphoton through
variables correlated with photon kinematics, isolation, and mass resolution.
To emulate the dominant kinematic effect of the CMS multivariate selection and to facilitate a comparison between the two
analyses, we impose (\ref{eq:50}) in our CMS-inspired implementation. As shown in Section~\ref{sec:features}, the characteristic momentum scale of the diphoton system arising from $h^0 \rightarrow A_h A_h$ naturally populates this boosted region. We have verified that varying the lower limit on $p_{\rm T}^{\gamma\gamma}$ within the range
$40 - 60~{\rm GeV} $ changes the signal efficiency only slightly, indicating that our conclusions are not sensitive to the precise choice of the $p_{\rm T}^{\gamma\gamma}$ threshold.

\renewcommand{\arraystretch}{1.35}
\begin{table}[t]
\centering
\small
\begin{tabular}{lcc}
\hline\hline
Selection criterion &
\hspace*{0.5cm} CMS~\cite{CMS:2026zsp} \hspace*{0.5cm}   &
ATLAS~\cite{ATLAS:2022abz} \\
\hline
$\ge 2$ reconstructed photons in the fiducial region
& 0.455 & 0.422 \\
Photon $p_{\rm T}$ threshold
& 0.205 & 0.166 \\
Diphoton mass window
& 0.181 & 0.149 \\
Isolation/reconstruction proxy ($\Delta R_{\gamma\gamma}$)
& 0.137 & 0.131 \\
Boosted diphoton selection
($p_{\rm T}^{\gamma\gamma}>50~{\rm GeV}$)
& 0.127 & 0.119 \\
\hline\hline
\end{tabular}
\caption{Cumulative efficiencies for $pp\rightarrow h^0 \rightarrow A_h A_h\rightarrow(\gamma\gamma)(gg)$
for an ALP mass  $M_A=13.6~\mathrm{GeV}$ at $\sqrt{s} = 13$ TeV.
CMS- and ATLAS-inspired selections are applied sequentially: {\it 1.} Photon 
reconstruction and fiducial acceptance (including a calorimeter crack veto); 
{\it 2.} Photon transverse-momentum requirements, (\ref{eq:pTgammaATL}) for ATLAS, (\ref{eq:pTgammaCMS}) and (\ref{eq:pTgammaCMS2}) for CMS;
{\it 3.} Diphoton invariant-mass window; 
{\it 4.} Photon-pair separation criterion (a proxy for photon isolation and reconstruction);
{\it 5.} Boosted-topology requirement (\ref{eq:50}). Final row gives the overall signal acceptance times efficiency
($A\times\epsilon$) obtained in our CMS- and ATLAS-inspired analyses.
}
\label{tab:cms_atlas_eff}
\end{table}

The detector-level cut flow is summarized in Table~\ref{tab:cms_atlas_eff}.
The resulting acceptance times efficiency is 12.7\% and 11.9\% for the CMS~\cite{CMS:2026zsp} and ATLAS~\cite{ATLAS:2022abz} analyses, respectively, for collisions at a center-of-mass energy of 13 TeV. The close agreement between the two analyses indicates that the signal kinematics for our model are well-matched to the phase-space region targeted by both experiments.

\subsubsection{Higgs parameter space consistent with the CMS excess}

Let us now compare the signal predictions of our model with the CMS and ATLAS results and determine the parameter range capable of generating an observable excess near $m_{\gamma\gamma}\simeq13.6$ GeV while remaining compatible with the limits of both experiments. This provides a quantitative assessment of whether a light ALP can simultaneously account for the observed CMS $3.5 \sigma$ excess and the ATLAS milder excess at that mass.

The Higgs boson branching fraction into undetected modes, ${\cal B}(h^0)_{\rm und}$, can in principle be large, as the effect of a larger width can be compensated by an overall increase of the couplings to all SM particles~\cite{Dobrescu:2012td}. A severe constraint on the ${\cal B}(h^0)_{\rm und}$ can be derived under the assumption that the Higgs coupling to the $W$ and $Z$ bosons are not larger than the SM values. This assumption, often denoted by $|\kappa_V| \leq 1$, is approximately valid as long as there are no $SU(2)_W$ triplets or higher representations that acquire VEVs~\cite{Dobrescu:2012td}.

The upper limit on the Higgs boson branching fraction into undetected modes set by ATLAS (Fig.~6 of~\cite{ATLAS:2022vkf}) is ${\cal B}(h^0)_{\rm und} < 12\%$ at  the $95\%$ CL limit, under the assumption of $|\kappa_V| \leq 1$.  CMS set less stringent limits (Fig.~14 of~\cite{CMS:2026nce}) on ${\cal B}(h^0)_{\rm und}$ for $|\kappa_V| \leq 1$: 22\% at the $2\sigma$ level, and 11\% at the $1\sigma$ level.   

\begin{figure}[t!]
\centerline{\hspace*{-.6cm}   \includegraphics[width=11.cm, angle=0]{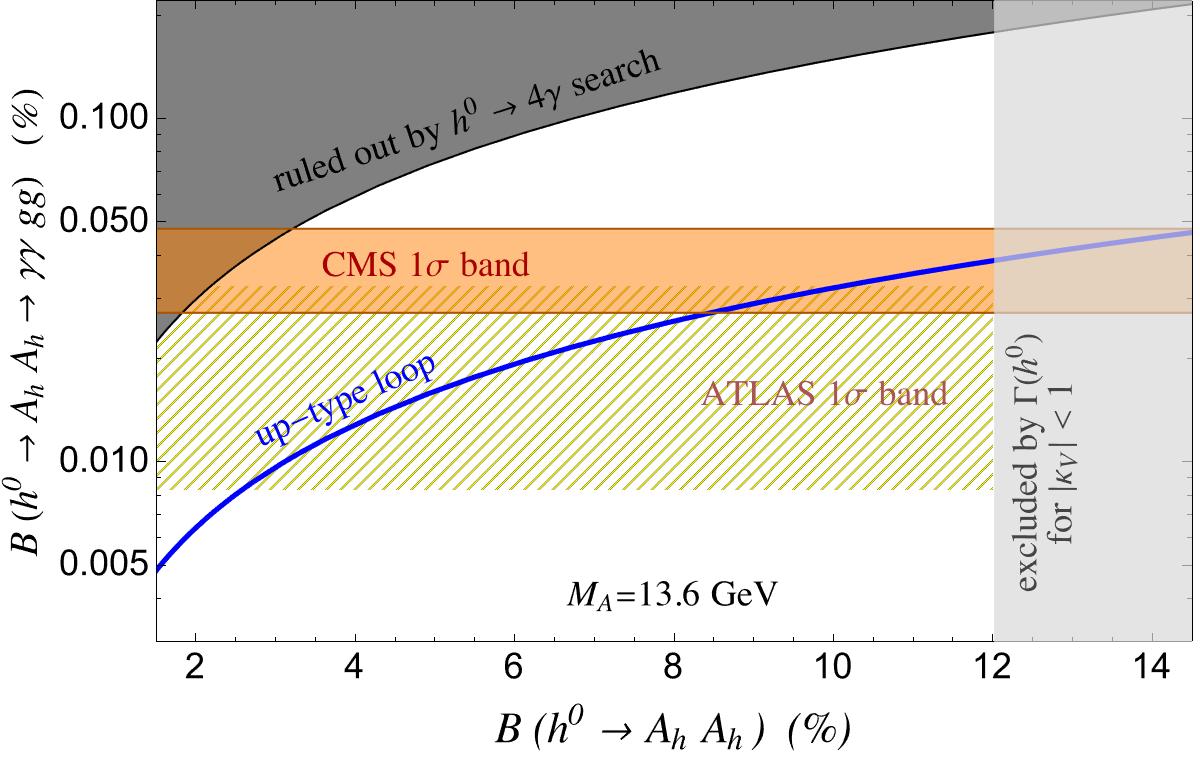}    } 
\vspace*{-.2cm}  
\caption{ Combined branching fraction for the Higgs ($h^0$) decay into $A_h A_h \to \gamma\gamma gg$ as a function of the 
branching fraction for $h^0 \to A_h A_h$. The solid blue curve represents the prediction of our model with charge-2/3 colored particles in the loop. The horizontal orange-shaded and yellow-hatched  bands are the $1\sigma$ regions allowed by the CMS~\cite{CMS:2026zsp} and ATLAS~\cite{ATLAS:2022abz} searches, respectively.  
\vspace*{2mm}
} 
\label{fig:hHphi}
\end{figure}

In Figure~\ref{fig:hHphi} we show the existing constraints on the Higgs branching fractions as well as the prediction of our model for $M_A = 13.6$ GeV.  
The horizontal axis represents the branching fraction of the Higgs boson into a pair of ALPs, 
$B(h^0 \to A_hA_h)$,  while the vertical axis represents the combined branching fraction of the full cascade decay:
\be
B(h^0 \to A_hA_h \to \gamma\gamma gg) = 2 B(h^0 \to A_hA_h) B(A_h \to \gamma\gamma) B(A_h \to gg)  ~~.
\ee 
Solid blue curve is the prediction for the $A_h$ decays induced by color-triplet particles of electric charge 2/3 running in the loop,
as shown for example in Figure~\ref{fig:u12loop}. Upper-side dark-gray shaded region is ruled out by the ATLAS~\cite{ATLAS:2023ian} and CMS~\cite{CMS:2025mwx} searches for $h^0 \to A_hA_h \to 4\gamma$. Right-side light-gray shaded region is ruled out by ATLAS~\cite{ATLAS:2022vkf} under the assumption of $|\kappa_V| \leq 1$. 

The $1\sigma$ region consistent with the CMS excess in the diphoton search~\cite{CMS:2026zsp} is the horizontal orange-shaded band in Figure~\ref{fig:hHphi}, while the analogous region for the ATLAS diphoton search is the horizontal yellow-striped band~\cite{ATLAS:2022abz}.
Remarkably, the two $1\sigma$ regions have some overlap, indicating that the ATLAS result is consistent with the CMS excess.

\begin{figure}[t]
    \centering
    \includegraphics[width=0.48\textwidth]{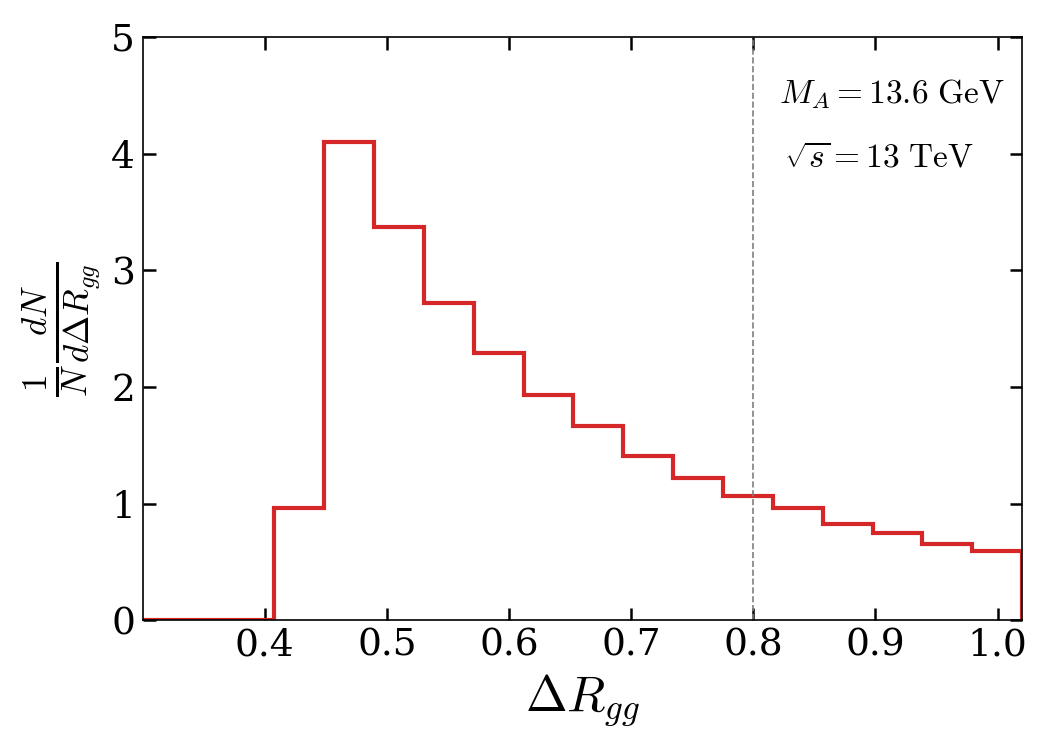}
    \hfill
    \includegraphics[width=0.496\textwidth]{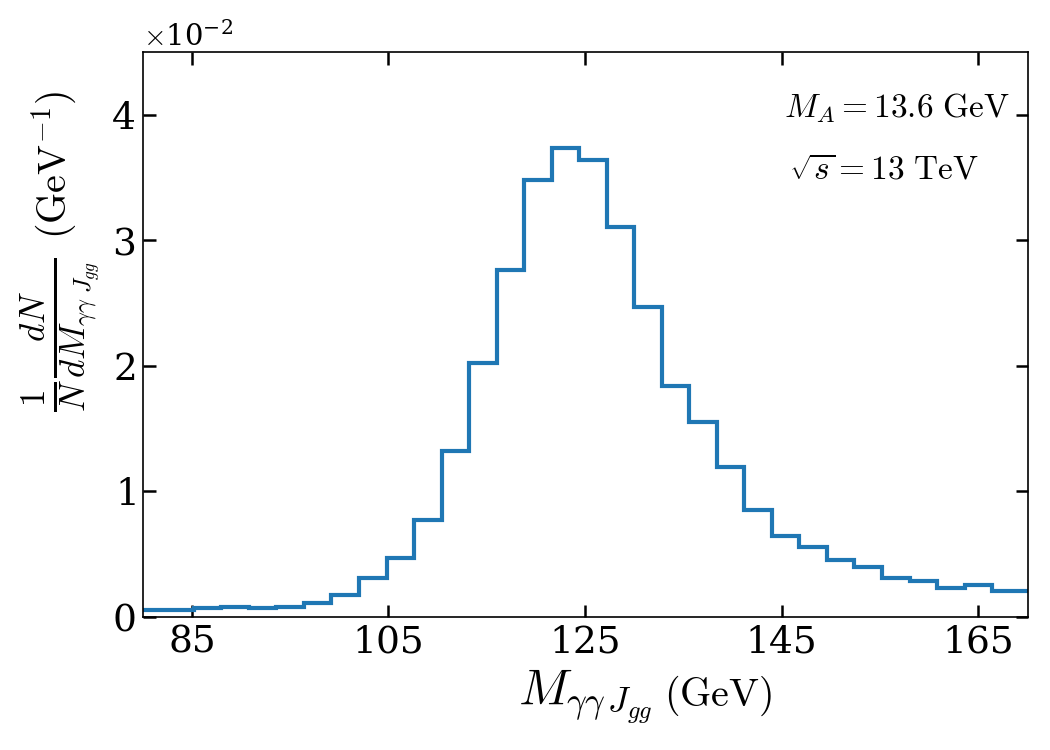}
    \caption{
    Kinematics of the $pp\to h^0\to A_hA_h\to(\gamma\gamma)(gg)$ signal for
    $M_A=13.6$~GeV and $\sqrt{s} = 13$ TeV.
    {\it Left:} normalized parton-level distribution of the angular separation
    $\Delta R_{gg}$ between the two gluons; for values to the left of the dashed line the two gluons form a single wide jet.
    {\it Right:} normalized reconstructed invariant-mass distribution
    $M_{\gamma\gamma J_{gg}}$, obtained by combining the two leading
    reconstructed photons with the groomed leading wide jet ($R = 0.8$) after requiring
    $\Delta R(J_{gg},\gamma_{1,2})>0.8$ and
    $11~{\rm GeV}<m_{\gamma\gamma}<16~{\rm GeV}$; 
    the peak near $125$~GeV reconstructs the Higgs boson. }
    \label{fig:signal_kinematics}
\end{figure}

\subsubsection{Additional features of the $(\gamma\gamma)(gg)$ signal}
\label{sec:gammasMore} 

Using the Monte Carlo setup described in Sec.~\ref{sec:analysis}, we can now verify the 
features of the gluon jets presented in Section~\ref{sec:features} beyond the simple scaling arguments discussed there.
The left-hand panel of Figure~\ref{fig:signal_kinematics} shows the
parton-level angular separation,  $ \Delta R_{gg}$, between the two gluons directly originating from the same
$A_h\to gg$ decay, for an ALP mass $M_A=13.6$~GeV.  The distribution exhibits the characteristic threshold and peak
structure expected from the 2-body decay of a boosted resonance,
with a sharp rise around $\Delta R_{gg}\simeq0.45$, in agreement with
the estimate in Eq.~(\ref{eq:deltaRestimate}), followed by a tail toward larger
separations due to the spread in the $A_h$ transverse momentum.  We find that approximately $62\%$
($74\%$) of the $A_h\to gg$ decays satisfy $\Delta R_{gg}<0.8$ ($1.0$).

Motivated by this collimation, we identify the hadronically decaying
$A_h$ with a large-radius jet $J_{gg}$. We reconstruct wide jets using
the anti-$k_T$ algorithm~\cite{Cacciari:2008gp} with radius parameter $R=0.8$.
The jets are subsequently groomed using the soft-drop
procedure~\cite{Larkoski:2014wba}, which removes soft, wide-angle radiation by
requiring the two branches encountered during declustering to satisfy
\begin{equation}
\frac{\min(p_{T1},p_{T2})}{p_{T1}+p_{T2}} > z_{\rm cut} \left(\frac{\Delta R_{12}}{R}\right)^\beta .
\end{equation}
We use $z_{\rm cut}=0.1$ and $\beta=0$, for which the grooming
condition is independent of the angular separation between the two
branches.  After parton showering, hadronization, and
detector simulation, the two highest-$p_{\rm T}$ reconstructed photons are
combined into a diphoton candidate, while the leading wide jet is
taken as the $J_{gg}$ candidate.  We require   $\Delta R(J_{gg},\gamma)>0.8$
for each of the two leading photons, to avoid overlap between the hadronic and electromagnetic objects,
and also require $11~{\rm GeV}<m_{\gamma\gamma}<16~{\rm GeV}$ to select diphoton
candidates consistent with the $A_h$ resonance.

The right-hand panel of Figure~\ref{fig:signal_kinematics} shows the invariant mass
$M_{\gamma\gamma J_{gg}}$ obtained by combining the diphoton system
with the soft-drop groomed 4-momentum of the leading wide jet.  The distribution exhibits a broad peak near $125$~GeV,
demonstrating that the mass scale of the parent Higgs boson can be
reconstructed from the $\gamma\gamma J_{gg}$ system.
Thus, in addition to the low-mass diphoton resonance, requiring consistency of the reconstructed $M_{\gamma\gamma J_{gg}}$ peak with the 
Higgs mass would further suppress the backgrounds.

Besides the signal with two gluons forming a wide jet, there is a substantial fraction of events where the two gluon jets can be resolved. Even though the background for these $\gamma\gamma j_g j_g$ events is expected to be larger (for a phenomenological study of this see \cite{Martin:2007dx}; an ATLAS search \cite{ATLAS:2018jnf} in this channel focused on $M_A \ge 20$ GeV), they can be included in a combination of searches that probe the $h^0 \rightarrow A_h A_h$ decays. 

We have focused here on Higgs production via gluon fusion. Searches for $h^0 \rightarrow A_h A_h \to (\gamma\gamma)(gg)$, with either merged or resolved jets, based on additional Higgs production mechanisms can also be performed, in ways related to those discussed in Section \ref{sec:features}. In particular,  associated $h^0$ production, similar to the right-hand diagram of Figure~\ref{fig:JJ}, would provide powerful methods of separating the signals from the backgrounds.

\bigskip

\section{Conclusions}  
\label{sec:conclusions}    \setcounter{equation}{0}

New particles at the 10 GeV scale are difficult to detect if they decay primarily to hadrons, due to the large QCD backgrounds at colliders. A type of particle that naturally can have mass of order 10 GeV or smaller is the ALP, because a shift symmetry protects its mass from quadratic divergences. 
ALPs have spin 0  and thus could couple to the SM through the Higgs portal (\ref{eq:AAHH}), in which case we label them by $A_h$.
 Furthermore, as the ALPs are generically expected to be much lighter than the SM Higgs boson, the latter  can decay into two $A_h$'s. 
 
 The global symmetry associated with the ALP forbids the Higgs portal coupling, but that symmetry must be explicitly broken to allow an ALP mass. Consequently, the small ALP mass implies a Higgs portal coupling much smaller than unity, so it is not surprising that the experimental searches at the LHC have not yet observed any nonstandard Higgs decays.
As follows from Eq.~(\ref{eq:BRh}), an effective portal coupling of $10^{-2}$ yields a branching fraction of 10\% for the Higgs boson decay into two $A_h$'s; that value is below the current constraints on non-detectable Higgs modes.

If $A_h$ couples to heavy colored particles, and it is not part of an weak doublet (so that decays into SM fermions~\cite{Dobrescu:2000yn}  are forbidden at tree level), then the decay to gluons $A_h \to gg$ has a branching fraction larger than 99\%. As a result, the Higgs boson can undergo the cascade decay $h^0 \to A_h A_h \to (gg)(gg)$ with a combined branching fraction of order 10\%. Currently, no search for this natural type of nonstandard Higgs decay has been performed. 

Although the QCD cross section for a 4-gluon final state is enormous, the topology involving $A_h$'s allows efficient background reduction without a large signal suppression. As each $A_h$ arising from $h^0$ decay are typically boosted, the two gluons form a wide jet ($J_{gg}$) with 2-prong substructure having a mass peak near the $A_h$ mass, $M_A$. An additional way of suppressing the background is to take advantage of the invariant-mass distribution of the system of two wide jets $J_{gg}$, which peaks near 125 GeV.
Moreover, a significant fraction of the Higgs bosons at the LHC, especially in the case of associated production,  are boosted so that the two $J_{gg}$ jets merge in a wider jet of (2+2)-prong substructure, again different from the QCD background. We thus urge the ATLAS and CMS Collaborations to search for this well-motivated Higgs decays into wide jets.

If the heavy colored particles are color triplets and carry electric charge 2/3, as in the case of up-type squarks (see the model presented in Section~\ref{sec:model}) or vectorlike quarks (see~\cite{Dobrescu:2000jt}), then the branching fraction of $A_h$ into photons is of order $10^{-3}$, and the mixed channel $h^0\to A_hA_h\to(\gamma\gamma)(gg)$ provides a promising ALP probe at the LHC. 
In Section~\ref{sec:gammas} we showed that this process can explain the CMS excess, of $3.5\sigma$ local significance, for a diphoton resonance at 13.6 GeV, and is simultaneously compatible with the small ($\sim \! 1.8\sigma$) ATLAS excess at the same mass.

Unlike a singly-produced diphoton resonance, the $h^0\to A_hA_h\to(\gamma\gamma)(gg)$ process includes distinctive hadronic activity (typically a wide jet $J_{gg}$) from the decay of the second $A_h$, which can be used by CMS and ATLAS to improve the sensitivity to $A_h$.  
Combining the wide jet with the diphoton system reconstructs a peak at the Higgs mass, so that requiring both a resonance in $m_{\gamma\gamma}$ near $M_A$ and a $\gamma\gamma J_{gg}$ resonance near $M_h$ would substantially reduce the background. Independently of the current CMS excess, the nonstandard Higgs decay signatures studied here provide well-motivated targets for 
new light particle searches at the Run~3 and High-Luminosity LHC, as well as at the FCC-ee and other future colliders.

\bigskip\bigskip\bigskip


{\bf Acknowledgments:} \ We would like to thank Yang Bai and KC Kong for insightful comments.
Fermilab is administered by Fermi Forward Discovery Group, LLC under Contract No. 89243024CSC000002 with the U.S. Department of Energy, Office of Science, Office of High Energy Physics.
SR is supported by the U.S.~Department of Energy under contracts No.\ DEAC02-06CH11357 at the Argonne National Laboratory.  SR would like to thank the University of Chicago, Fermilab and Perimeter Institute where a significant part of this work was carried out.
\smallskip

\smallskip



\begin{thebibliography}{99} 

\bibitem{Bauer:2017ris}
M.~Bauer, M.~Neubert and A.~Thamm,
``Collider Probes of Axion-Like Particles,''
JHEP \textbf{12}, 044 (2017)
[arXiv:1708.00443 [hep-ph]].\\
S.~Knapen, S.~Kumar and D.~Redigolo,
``Searching for axionlike particles with data scouting at ATLAS and CMS,''
Phys. Rev. D \textbf{105}, no.11, 115012 (2022)
[arXiv:2112.07720 [hep-ph]].\\
G.~Alonso-{\'A}lvarez, J.~Jaeckel and D.~D.~Lopes,
``Tracking axion-like particles at the LHC,''
[arXiv:2302.12262 [hep-ph]].\\
A.~Biek{\"o}tter and K.~Mimasu,
``Axions and Axion-like particles: collider searches,''
[arXiv:2508.19358 [hep-ph]].

\bibitem{CMS:2016ltu}
V.~Khachatryan \textit{et al.} [CMS],
``Search for narrow resonances in dijet final states at $\sqrt{s}=$ 8 TeV with the novel CMS technique of data scouting,''
Phys. Rev. Lett. \textbf{117}, no.3, 031802 (2016)
[arXiv:1604.08907 [hep-ex]].

\bibitem{CMS:2024zhe}
A.~Hayrapetyan \textit{et al.} [CMS],
``Enriching the physics program of the CMS experiment via data scouting and data parking,''
Phys. Rept. \textbf{1115}, 678-772 (2025)
[arXiv:2403.16134 [hep-ex]].

\bibitem{Dobrescu:2000jt}
B.~A.~Dobrescu, G.~L.~Landsberg and K.~T.~Matchev,
``Higgs boson decays to CP odd scalars at the Tevatron and beyond,''
Phys. Rev. D \textbf{63}, 075003 (2001)
[arXiv:hep-ph/0005308].

\bibitem{Martin:2007dx}
A.~Martin, ``Higgs cascade decays to gamma gamma + jet jet at the LHC,''
[arXiv:hep-ph/0703247].

\bibitem{CMS:2025mwx}
 CMS Collaboration,
report PAS-EXO-24-025, July 2025. \\
A.~Hayrapetyan \textit{et al.} [CMS],
``Search for exotic Higgs boson decays $H\to\mathcal{AA}$ with $\mathcal{AA} \to \gamma\gamma$ in events with a semi-merged topology in proton-proton collisions at $\sqrt{s}$ = 13 TeV,''
[arXiv:2601.00183 [hep-ex]].

\bibitem{ATLAS:2023ian}
G.~Aad \textit{et al.} [ATLAS],
``Search for short- and long-lived axion-like particles in $H\rightarrow a a \rightarrow 4\gamma $ decays with the ATLAS experiment at the LHC,''
Eur. Phys. J. C \textbf{84}, no.7, 742 (2024)
[arXiv:2312.03306 [hep-ex]].    


\bibitem{Chang:2006bw}
S.~Chang, P.~J.~Fox and N.~Weiner,
``Visible cascade higgs decays to four photons at hadron colliders,''
Phys. Rev. Lett. \textbf{98}, 111802 (2007)
[arXiv:hep-ph/0608310].

\bibitem{Brivio:2026qtx}
I.~Brivio, S.~Meoni and D.~Pagani,
``ALP pair production at the LHC,''
[arXiv:2607.21712 [hep-ph]].


\bibitem{CMS:2026zsp}
A.~Hayrapetyan \textit{et al.} [CMS],
``Search for a narrow resonance with a mass between 10 and 70 GeV decaying to a pair of photons in proton-proton collisions at $\sqrt{s} = 13$ TeV,'' [arXiv:2603.03250 [hep-ex]].

\bibitem{ATLAS:2022abz}
G.~Aad \textit{et al.} [ATLAS],
``Search for boosted diphoton resonances in the 10 to 70 GeV mass range using 138 fb$^{-1}$ of 13 TeV $pp$ collisions,''
JHEP \textbf{07}, 155 (2023)
[arXiv:2211.04172 [hep-ex]].

\bibitem{Dobrescu:1999gv}
B.~A.~Dobrescu,
``Minimal composite Higgs model with light bosons,''
Phys. Rev. D \textbf{63}, 015004 (2001)
[arXiv:hep-ph/9908391].

\bibitem{Dobrescu:2000yn}
B.~A.~Dobrescu and K.~T.~Matchev,
``Light axion within the next-to-minimal supersymmetric standard model,''
JHEP \textbf{09}, 031 (2000)
[arXiv:hep-ph/0008192].

\bibitem{Dermisek:2006wr}
R.~Dermisek and J.~F.~Gunion,
``The NMSSM close to the R-symmetry limit and naturalness in $h \to aa$ decays for $m_a < 2m_b$,''
Phys. Rev. D \textbf{75}, 075019 (2007)
[arXiv:hep-ph/0611142].

\bibitem{Datta:2022bvg}
A.~Datta, M.~Guchait, A.~Roy and S.~Roy,
``Hunting ewinos and a light scalar of Z$_{3}$-NMSSM with a bino-like dark matter in top squark decays at the LHC,''
JHEP \textbf{11}, 081 (2023)
[arXiv:2211.05905 [hep-ph]].

\bibitem{Bernreuther:2023uxh}
E.~Bernreuther and B.~A.~Dobrescu,
``Vectorlike leptons and long-lived bosons at the LHC,''
JHEP \textbf{07}, 079 (2023)
[arXiv:2304.08509 [hep-ph]]. \\
M.~R.~Buckley, D.~Shih and I.~R.~Wang,
``Hiding in the shadow of the Upsilon: ditaus from a light pseudoscalar,''
[arXiv:2605.29289 [hep-ph]]. \\
C.~Houghton, A.~Lath, J.~Reichert and S.~Thomas,
``Signals of new resonances from di-lepton non-universality in the bottomonium mass region at the Large Hadron Collider,''
[arXiv:2605.31357 [hep-ph]].

\bibitem{ATLAS:2024vpj}
G.~Aad \textit{et al.} [ATLAS],
``Search for decays of the Higgs boson into a pair of pseudoscalar particles decaying into $bb\tau^+ \tau^-$ using pp collisions at $\sqrt{s}=13$ TeV,''  Phys. Rev. D \textbf{110}, no.5, 052013 (2024)
[arXiv:2407.01335 [hep-ex]].

\bibitem{CMS:2026mwx}
A.~Hayrapetyan \textit{et al.} [CMS],
``Search for low-mass resonances decaying to $\tau\tau$ and measurement of the $\Upsilon \to \tau\tau$ decay in proton-proton collisions at $\sqrt{s}$ = 13.6 TeV,''  [arXiv:2605.25103 [hep-ex]]. \\
A.~Hayrapetyan \textit{et al.} [CMS],
``Search for the decay of the Higgs boson to a pair of light pseudoscalar bosons in the final state with four bottom quarks in proton-proton collisions at $ \sqrt{\textrm{s}} $ = 13 TeV,''  JHEP \textbf{06}, 097 (2024)
[arXiv:2403.10341 [hep-ex]].

\bibitem{ParticleDataGroup:2024cfk}
S.~Navas \textit{et al.} [Particle Data Group],
``Review of particle physics,''
Phys. Rev. D \textbf{110}, no.3, 030001 (2024)

\bibitem{Djouadi:1998az}
A.~Djouadi,
``Squark effects on Higgs boson production and decay at the LHC,''
Phys. Lett. B \textbf{435}, 101-108 (1998)
[arXiv:hep-ph/9806315 [hep-ph]].


\bibitem{Anastasiou:2016cez}
C.~Anastasiou, C.~Duhr, F.~Dulat, E.~Furlan, T.~Gehrmann, F.~Herzog, A.~Lazopoulos and B.~Mistlberger,
``High precision determination of the gluon fusion Higgs boson cross-section at the LHC,''
JHEP \textbf{05}, 058 (2016)
[arXiv:1602.00695 [hep-ph]].    


\bibitem{Alloul:2013bka}
A.~Alloul, N.~D.~Christensen, C.~Degrande, C.~Duhr and B.~Fuks,
``FeynRules  2.0 - A complete toolbox for tree-level phenomenology,''
Comput. Phys. Commun. \textbf{185}, 2250-2300 (2014)
[arXiv:1310.1921 [hep-ph]].

\bibitem{Darme:2023jdn}
L.~Darm{\'e}, C.~Degrande, C.~Duhr, B.~Fuks, M.~Goodsell, G.~Heinrich, V.~Hirschi, S.~H{\"o}che, M.~H{\"o}fer and J.~Isaacson, 
``UFO 2.0: the {\textquoteleft}Universal Feynman Output{\textquoteright} format,''
Eur. Phys. J. C \textbf{83} (2023) no.7, 631
[arXiv:2304.09883 [hep-ph]].

\bibitem{Alwall:2014hca}
J.~Alwall \textit{et al.}, 
``The automated computation of tree-level and next-to-leading order differential cross sections, and their matching to parton shower simulations,''
JHEP \textbf{07}, 079 (2014)
[arXiv:1405.0301 [hep-ph]].

\bibitem{Ball:2013hta}
R.~D.~Ball \textit{et al.} [NNPDF],
``Parton distributions with QED corrections,''
Nucl. Phys. B \textbf{877}, 290-320 (2013)
[arXiv:1308.0598 [hep-ph]].  

\bibitem{Sjostrand:2007gs}
T.~Sjostrand, S.~Mrenna and P.~Z.~Skands,
``A Brief Introduction to PYTHIA 8.1,''
Comput. Phys. Commun. \textbf{178} (2008), 852-867
[arXiv:0710.3820 [hep-ph]].

\bibitem{deFavereau:2013fsa}
J.~de Favereau \textit{et al.} [DELPHES 3],
``DELPHES 3, A modular framework for fast simulation of a generic collider experiment,''
JHEP \textbf{02} (2014), 057
[arXiv:1307.6346].

\bibitem{Brun:1997pa}
R.~Brun and F.~Rademakers,
``ROOT {\textemdash} An object oriented data analysis framework,''
Nucl. Instrum. Meth. A \textbf{389}, no.1-2, 81-86 (1997)


\bibitem{Dobrescu:2012td}
B.~A.~Dobrescu and J.~D.~Lykken,
``Coupling spans of the Higgs-like boson,''
JHEP \textbf{02}, 073 (2013)
[arXiv:1210.3342 [hep-ph]].

\bibitem{ATLAS:2022vkf}
G.~Aad \textit{et al.} [ATLAS],
``A detailed map of Higgs boson interactions by the ATLAS experiment ten years after the discovery,''
Nature \textbf{607}, no.7917, 52-59 (2022)
[erratum: Nature \textbf{612}, no.7941, E24 (2022)]
[arXiv:2207.00092 [hep-ex]].
    
\bibitem{CMS:2026nce}
A.~Hayrapetyan \textit{et al.} [CMS],
``Combined measurements and interpretations of Higgs boson production and decay in proton-proton collisions at $\sqrt{s}$ = 13 TeV,''
[arXiv:2602.18611 [hep-ex]].    

\bibitem{Cacciari:2008gp}
M.~Cacciari, G.~P.~Salam and G.~Soyez,
``The anti-$k_t$ jet clustering algorithm,''
JHEP \textbf{04}, 063 (2008)
[arXiv:0802.1189 [hep-ph]].

\bibitem{Larkoski:2014wba}
A.~J.~Larkoski, S.~Marzani, G.~Soyez and J.~Thaler,
``Soft Drop,''
JHEP \textbf{05}, 146 (2014)
[arXiv:1402.2657 [hep-ph]].
    
\bibitem{ATLAS:2018jnf}
M.~Aaboud \textit{et al.} [ATLAS],
``Search for Higgs boson decays into pairs of light (pseudo)scalar particles in the $\gamma\gamma jj$ final state in $pp$ collisions at $\sqrt{s}=13$ TeV,''
Phys. Lett. B \textbf{782}, 750-767 (2018)
[arXiv:1803.11145 [hep-ex]].    
    
\end{thebibliography}
\end{document}